\documentclass [12pt,elsart,epsfig]{article}     
\usepackage{epsfig}
\begin{document}
\begin {center}
{\Large \bf The $f_2(1565)$ in $\bar pp \to (\omega \omega )\pi ^0$ interactions 
at rest}
\vskip 5mm

{C.A. Baker$^c$, B.M. Barnett$^b$, C.J. Batty$^c$, K. Braune $^f$, D.V. Bugg$^e$, 
O. Cramer$^f$, V. Cred\' e$^a$, N. Djaoshvili$^d$, 
W. D\" unnweber$^f$, M.A. Faessler$^f$, N.P. Hessey$^f$,
P. Hidas$^b$,  C. Hodd$^e$, D. Jamnik$^f$, H. Kilinowsky$^a$, 
J. Kisiel$^d$, E. Klempt$^a$, C. Kolo$^f$, L. Montanet$^d$, B. Pick$^a$, 
W. Roethel$^f$, A. Sarantsev$^g$, I. Scott$^e$, C. Stra$\beta$burger$^a$,
U. Thoma$^a$, C. V\" olcker$^f$, S. Wallis$^f$, D. Walther$^f$, K. Wittmack$^a$, 
B.S.~Zou$^{b}$ \footnote{Now at IHEP, Beijing 100039, China} \\
{\normalsize $^a$ \it Univsersit\" at Bonn, D-53115 Bonn, Germany}\\
{\normalsize $^b$ \it Academy of Science, H-1525 Budapest, Hungary}\\
{\normalsize $^c$ \it Rutherford Appleton Laboratory, Chilton, Didcot OX11 0QX,UK}\\
{\normalsize $^d$ \it CERN, CH-1211 Geneva 4, Switzerland}\\
{\normalsize $^e$ \it Queen Mary and Westfield College, London E1\,4NS, UK}\\
{\normalsize $^f$ \it Universit\" at M\" unchen, D-80333 M\" unchen, Germany}\\ 
{\normalsize $^g$ \it PNPI, Gatchina, St. Petersburg district, 188350, Russia}\\

[3mm]}
\end {center}

\begin{abstract}
Data are presented on the reaction $\bar pp \to \omega \omega \pi ^0$ at 
rest from the Crystal Barrel detector.
These data identify a strong signal due to $f_2(1565) \to \omega \omega$.
The relative production from initial $\bar pp$ states $^3P_2$, $^3P_1$ and
$^1S_0$ is well determined from $\omega \omega$ decay angular correlations;
P-state annihilation dominates strongly.
A combined fit is made with data on $\bar pp \to 3\pi ^0$ at rest, where
$f_2(1565) \to \pi ^0 \pi ^0$ is observed.
A Flatt\' e formula is fitted to the $f_2(1565)$, including the $s$-dependence
of decay widths to $\omega \omega$ and $\rho \rho$.
The data then determine the K-matrix mass, $M = 1598 \pm 11(stat) \pm 9(syst)$
 MeV.
The decay width to $2\pi$ is very small, of order 2\% of the total width.
\end{abstract}

We investigate the $f_2(1565)$ through its decays to
$\omega \omega$ in the process
$\bar pp \to \omega \omega \pi ^0$ at rest.
Data on $\bar pp \to 3\pi ^0$ at rest are also used to estimate
relative widths to $\pi \pi$ and $\omega \omega$.

The first clear identification of the $f_2(1565)$ was by May et al.
in the Asterix experiment on $\bar pp \to \pi ^-\pi ^+\pi ^0$ at rest [1,2].
Subsequently, it has been identified by the Obelix collaboration in
$\bar np \to \pi ^-\pi ^+\pi ^+$ [3], and in Crystal Barrel data on
$\bar pp \to 3\pi ^0$ at rest [4,5].
However, it has become clear that the $f_2(1565)$ also couples strongly
to $\omega \omega$ (and therefore probably to $\rho \rho$).
Abele et al. [6] found a strong cusp in the $\pi \pi$ D-wave
at the $\omega \omega$ threshold in the $3\pi ^0$ data.
This led to the identification of $f_2(1565)$ as the same $2^+$ resonance as
observed  by GAMS [7] and VES [8]
in $\omega \omega$ just above threshold.
Their results are listed under $f_2(1640)$ by the Particle Data Group (PDG)
[9], who also list other sightings there and under $f_2(1565)$.
The properties of this resonance have remained elusive.
Here, we show that it is important to fit with a Flatt\' e formula which
includes the $s$-dependence of decays
to $\pi \pi$, $\rho \rho$ and $\omega \omega$; there is a sharp cusp at 1564
MeV due to the opening of the $\omega \omega$ channel.

In outline, the elements in the analysis are as follows.
Firstly, $\omega \omega$ correlations in the Crystal Barrel data
on the $\omega \omega \pi^0$ channel
determine well the relative amounts of annihilation from
initial states $^1S_0$, $^3P_1$ and $^3P_2$.
Secondly, earlier Crystal Barrel data on the
$3\pi ^0$ channel determine well the lower edge of the resonance.
Both of these sets of data are readily fitted with a Breit-Wigner amplitude
which includes the $s$-dependence of widths for decays to $\rho \rho$,
$\omega \omega$ and $\pi \pi$.
However, the upper side of the resonance is obscured in Crystal Barrel
data by the centrifugal barriers associated with the production
reaction; there is an $L = 2$ barrier for production from $^1S_0$ and
$L = 1$ barriers for production from $^3P_1$ and $^3P_2$. These
barriers cut off the resonance on its upper side in $\bar pp$ data at rest.
The data from GAMS and VES for the $\omega \omega$ channel provide a
good determination of the upper side of the resonance. We include the
VES data of Ref. [9] into the analysis in order to provide this constraint.

Firstly, we present technical aspects of data selection.
The new $\omega \omega \pi ^0$ data reported here concern events where both
$\omega$ decay to $\pi ^+\pi ^-\pi ^0$, so the final state is
$2\pi ^+ 2\pi ^- 3\pi ^0$. They were taken with the
Crystal Barrel detector [10]
using a $\bar p$ beam of 200 MeV/c stopping in liquid hydrogen.
In early runs, the target was surrounded by two cylindrical
multiwire chambers for triggering, followed by a cylindrical jet drift
chamber measuring charged particles. In later runs, the multiwire chambers
were replaced by a silicon vertex detector.
The outermost layers of the jet drift chamber covered $\sim 70\%$ of the
solid angle.
Surrounding this drift chamber was a barrel of 1380 CsI crystals
detecting photons with good resolution and high efficiency over a solid
angle 98\% of $4\pi $.
A solenoidal magnet provided a field of 1.5T.

The present data sample consists of $9.4$M triggered 4-prong events.
The trigger requires 4 hits in the multiwire chambers
(or Si vertex detector);
events with 4 long tracks are selected by demanding a hit in
the first 3 and last 4 layers of the jet drift chamber.
Off-line analysis selects $\sim 2.4$M events with four well reconstructed
tracks and balancing charges.
Gamma rays are selected by demanding a shower with energy larger than 10 MeV.
Any energy deposits matching up with charged tracks are rejected.
To eliminate the risk of shower energy being lost down the beam-pipe,
events are rejected
if the centroid of any shower lies in crystals
immediately adjoining the entrance or exit beam-pipe.
These cuts reduce the sample to $\sim 1.3$M.

Events with 4 charged tracks and $\ge 6\gamma$ are then submitted to a 7C fit
to $2\pi ^+ 2\pi ^- 3\pi ^0$;
11,679 events survive with a confidence level above $0.1\%$.
Almost all come from events containing only 6 or 7 showers.
In the latter case, one shower is interpreted as arising from a
so-called `split-off' in the CsI crystals.
This is caused by nuclear interaction of a charged particle,
which generates neutrons that undergo secondary interactions in nearby 
crystals;
those low energy deposits close to charged tracks are  discarded, but
some neutrons convert far from parent tracks.

\begin{figure}
\begin{center}
\epsfig{file=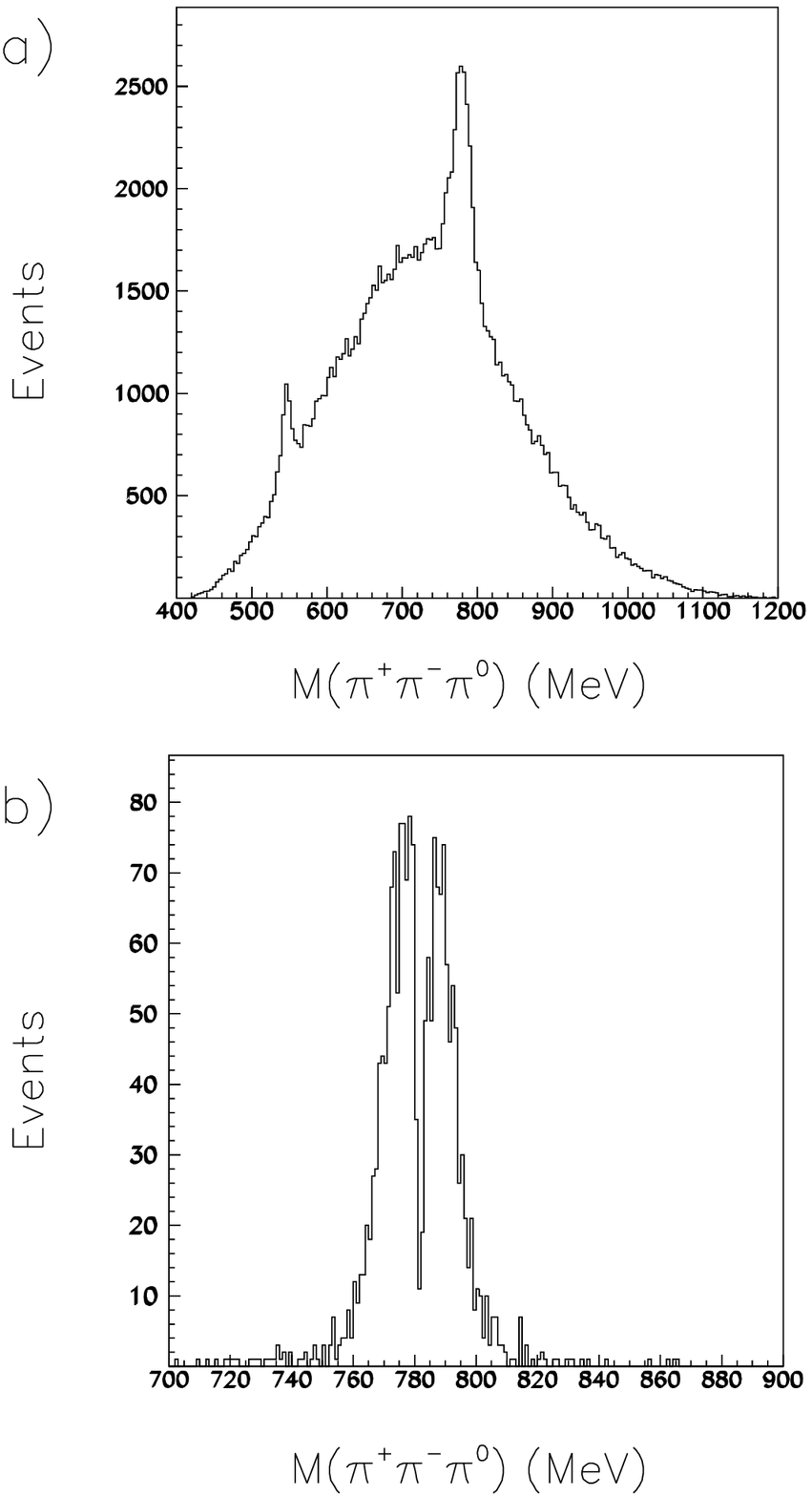,width=9.9cm}\
\vskip -18.3cm
\hskip 0.24cm
\epsfig{file=FIGURE10.EPS,width=9.9cm}\
~\
\caption{(a) The $\pi ^+\pi ^-\pi ^0$ mass spectrum from events fitting the
$2\pi ^+2\pi ^-3\pi ^0$ hypothesis with confidence level above $0.1\%$; 
(b) the mass distribution of the second best $\omega$ combination.}
\end{center}
\end{figure}


Turning now to physics results,
we consider first the branching ratio to $\omega \omega \pi ^0$.
The Monte Carlo simulation is unable to provide a precise efficiency for
the reconstruction of charged tracks, due to the `split-offs'.
Accordingly, we normalise the
$\omega \omega \pi ^0$ branching ratio to that for
$\omega \eta \pi ^0$. The latter is known from an earlier study of
$7\gamma$ events [11] and is $(6.8 \pm 0.5) \times 10^{-3}$.
Fig. 1(a) illustrates the mass spectrum of $\pi ^+\pi ^- \pi ^0$ combinations.
We estimate $930 \pm 50 $ $\eta \omega \pi ^0$ events and
$2000 \pm 50$ $\omega \omega \pi ^0$.
Correcting for the branching ratios $BR[\omega \to \pi ^+\pi ^0 \pi ^0]
= 0.888 \pm 0.007$ and $BR[\eta \to \pi ^+\pi ^0 \pi ^0]
= 0.274 \pm 0.026$, we find a branching fraction
\begin {equation}
BR[\bar pp \to \omega \omega \pi ^0] = (4.5 \pm 0.7) \times 10^{-3}.
\end {equation}

We next consider combinatorics and backgrounds.
The kinematic fit to $\omega \omega \pi ^0$ has 12 different combinations of
pions (and 180 permutations including individual photons).
It is possible that in some cases the confidence level of an
incorrect combination will be higher than that of the correct combination.
To reduce this problem, two further cuts have been applied, based on Monte
Carlo simulations.
The first is that there should be two or less combinations above 5\% confidence
level. The second
is that the ratio of the confidence level of the best combination to that of
the second best should be above $0.6$.
With these cuts, the Monte Carlo simulation estimates that wrong
combinations are less than $16\%$. They occur mostly amongst
similar geometries, which the simulations show to have small effects on
the fitted physics.

A possible source of background to $\omega \omega \pi ^0$ is the
$\omega \pi ^+\pi ^- 2\pi ^0$ channel.
To investigate this, we form all possible combinations of $\pi ^+\pi ^- \pi
^0$, select the best $\omega$ and then plot the invariant mass
distribution of the second best combination.
This is shown in Fig. 1(b).
There is a dip at the $\omega$ mass, due to selection of the best one.
The second best combination peaks at the $\omega$, 
but there is a background which,
extrapolated under the $\omega$, amounts to $\sim 4\%$.
A cut
is applied such that the invariant mass of the second best $\omega $ combination
lies between 752 and 812 MeV, so as to reject obvious background events.
Finally a kinematic fit is made to $\omega \omega \pi ^0$, and
events fitting with confidence level above $5\%$ are accepted.
There are eventually 1346 accepted events and 4677 from the Monte Carlo 
simulation.

\begin{figure}
\begin{center}
\epsfig{file=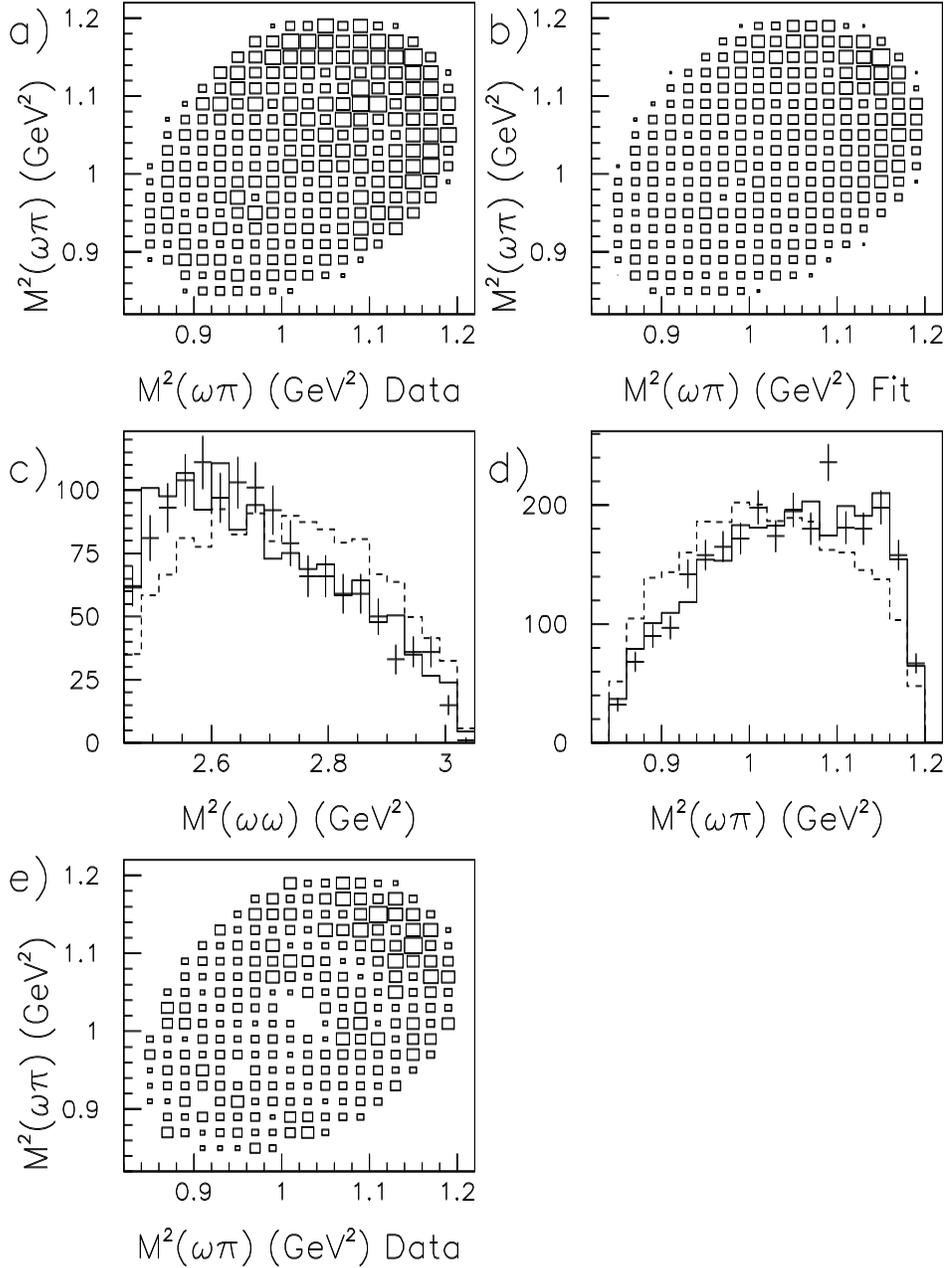,width=14cm}\
\vskip -18.687cm
\epsfig{file=TOTALPLO.EPS,width=14cm}\

\vskip -1mm
\caption{Dalitz plots for (a) data, (b) fit. Projections on to 
(c) $M^2(\omega \pi )$ and $M^2(\omega \omega)$; 
points with errors are data, full histograms show the fit and dashed 
histograms show phase space after acceptance cuts; 
(e) the Dalitz plot for $8\gamma$ events $\bar pp \to \omega \omega \pi ^0$ 
in hydrogen gas.}
\end{center}
\end{figure}

Fig. 2 (a) shows the Dalitz plot for data and Figs. 2(c) and (d) the mass
projections for $\omega \pi$ and $\omega \omega$ combinations.
Full histograms show the result of the maximum likelihood fit described
below; dashed histograms show distributions derived from phase space
after acceptance cuts. The Dalitz
plot from the fit is shown in Fig. 2(b).
There is a clear peaking of events towards the top right-hand edge of the
Dalitz plot. This could be due to $f_2(1565) \to \omega \omega$,
$f_0(1500) \to \omega \omega$ or $b_1(1235)\omega$.
Distinctions may be made using
angular correlations for decay of the two $\omega$.

In addition, data are also available for the $\omega \omega \pi ^0$ final state
from $8\gamma$ data, where both $\omega \to \pi ^0 \gamma$.
The Dalitz plot for 740 events from annihilation in hydrogen gas
is shown in Fig. 2(e). The $f_2(1565)$ appears prominently near
threshold; it is enhanced in P-state annihilation in gas.
These data have been shown to give amplitudes consistent with the
analysis presented here.
However, we shall not present their analysis in full.
The reason is that much of
the spin information about $\omega \omega$ correlations is lost
in the polarisations carried by decay photons from the $\omega$ decays.
In this respect, charged decays of the $\omega$ are greatly superior, since
no spin information is lost.

The primary processes which we consider in the amplitude analysis
are:
\begin {eqnarray}
^3P_2,~^3P_1,~^1S_0~~\bar pp &\to & f_2(1565)\pi \\
              ^1S_0~~\bar pp &\to & f_0(1500)\pi \\
              ^1S_0~~\bar pp &\to & f_0(1770)\pi \\
              ^1S_0~~\bar pp &\to & \sigma \pi \\
              ^1S_0~~\bar pp &\to &b_1(1235)\omega .
\end {eqnarray}
Here $\sigma$ denotes a slowly varying $0^+$ component decaying to
$\omega \omega$,
parametrised as a constant amplitude.
We have also tried contributions for production of $\omega \omega$
with $J^P = 0^-$, $1^-$ or $2^-$ resonances having widths set to 250 MeV,
but find them to be consistent with zero within errors.

\begin{figure}
\begin{center}
\vskip -6mm
\epsfig{file=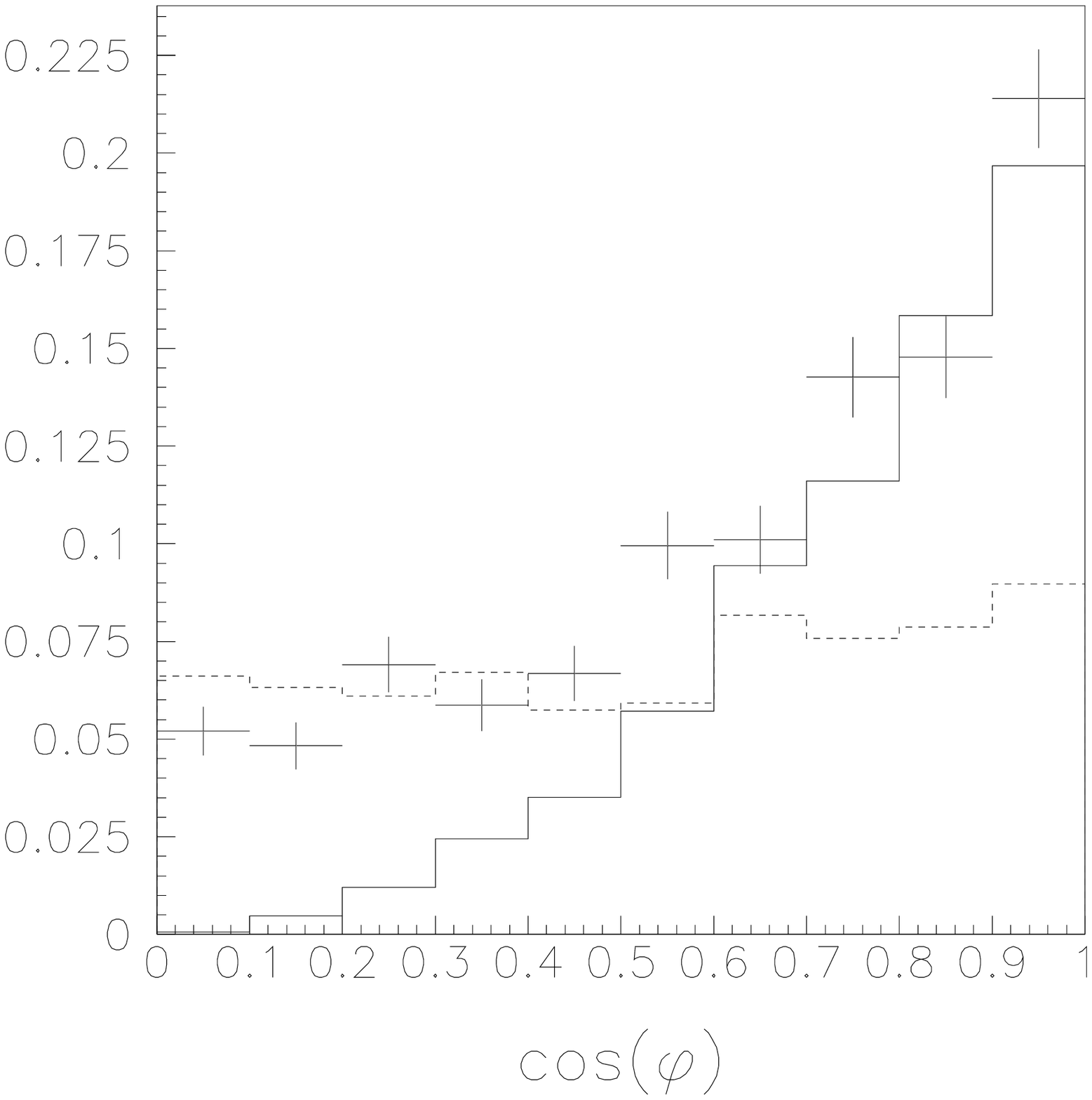,width=8cm}\
\vskip -8.025cm
\epsfig{file=N1N2ALL2.EPS,width=8cm}\
\vskip -1mm
\caption{The $\cos \phi$ distribution for data (points with errors), 
$0^+$ (full histogram) and $2^+$ (dashed histogram).}
\end{center}
\end{figure}

For spin 0 resonances, reactions (3)--(5), the amplitude is
proportional to $n_A.n_B$, where $n = p_1 \wedge p_2$ and $p_{1,2}$
are 3-momenta of $\pi ^+$ and $\pi ^-$ from decays of each $\omega _{A,B}$
in their rest frames.
For spin 2, amplitudes are given by complicated tensor expressions [12]
in terms of $n_A$, $n_B$ and the decay angles $\theta _i$ of each resonance to
$\omega \omega$ in its rest frame.
Fig. 3 illustrates the projection of decay probabilities on to
$\cos \phi$, where $\phi$ is the angle between the vectors
$n_A$ and $n_B$ normal to the $\omega$ decay planes. For spin 0, the
distribution is simply $\cos ^2 \phi$; for spin 2, it is
much flatter in terms of $\phi$, but in addition has a distinctive
correlation between $\phi$, $\theta _1$ and $\theta _2$ which is
included in the amplitude analysis.
The data clearly demand a mixture of $0^+$ and $2^+$.

As a check, we have examined four slices of $\omega \omega$ mass and determined
the intensities of the $0^+$ signal and $2^+$ originating from
$^3P_2$, $^3P_1$ and $^1S_0$ in each slice.
The $2^+$ signal peaks strongly near the $\omega \omega$ threshold,
while the $0^+$ signal extends fairly uniformly across the whole
$\omega \omega$ mass range.
It is therefore clear that the $f_2(1565)$ accounts for a large
proportion of the observed events.
This is also evident from Fig. 2(e).

Some further simplifications in the analysis are possible.
We find that there is little distinction between $\sigma \pi$ and
$f_0(1770)\pi$ final states, because of the limited $\omega \omega$
mass range.
We therefore drop $f_0(1770)$.
Secondly, there is little distinction between the $f_0(1500)\pi$ channel
and $b_1(1235)\omega$. The reason is that both processes
involve mostly orbital angular momentum $L = 0$ in the production reaction 
and both enhance the upper right-hand  edge of the Dalitz plot.
[In fitting $b_1(1235)$, we allow for decays to $\omega \pi$ with both 
$\ell = 0$ and 2 and take the small (0.29) D/S ratio of amplitudes from 
the PDG value].
The $b_1(1235)\omega$ channel
gives a somewhat better log likelihood than $f_0(1500)$.
However, if both processes
are introduced, they fit to large magnitudes
with an unreasonably large destructive interference between them.
It is therefore necessary to make a choice between them.
We now argue from other data that the contribution from $f_0(1500)$ is small.

Firstly, in amplitude analyses of $\bar pp \to 3\pi ^0$ at rest [4--6],
there is no indication of any cusp in the $f_0(1500)$ amplitude
at the $\omega \omega$ threshold.
Secondly, data on $J/\Psi \to \gamma (4\pi)$ require
$f_0(1500)$ decaying dominantly into $\sigma \sigma$ rather than
$\rho \rho$ [13]; at the quark level, coupling to $\rho \rho$ is
three times that to $\omega \omega$, making the $\omega \omega$ decay
weak. Thirdly, analysis of Crystal Barrel data at rest
to $5\pi$ shows a large signal for $f_0(1500) \to \sigma \sigma$,
but little or none for decay to $\rho \rho$ [14].
From these sources, we estimate that the branching ratio
of $f_0(1500)$ to $\omega \omega$ is less than $4\%$ of that to $\pi \pi$.
The branching fraction for $\bar pp \to f_0(1500)\pi$ followed by
the decay $f_0(1500) \to \pi \pi $ is known from $3\pi ^0$ data
to be $\simeq 2.45 \times 10^{-3}$ at rest [6]; hence the branching fraction
for its decay via $\omega \omega$ is below $10^{-4}$.
Using equn. (1), $f_0(1500) \to \omega \omega$
must be less than $2.1\%$ in intensity in present data
and we drop it from the analysis.
Including it with an intensity 3 times this limit produces only a small
interference with the $b_1(1235)\omega$ final state and has little effect on
conclusions concerning $f_2(1565)$.

Contributions to the fit are shown in Table 1.
Those from $\sigma \pi$ and $b_1(1235)\omega$ are lumped together
because of interferences which are irrelevant to the present
discussion.
It is obvious that P-state annihilation dominates
production of the $f_2(1565)$. The ratio of intensities from
$^3P_2$ and $^3P_1$ is close to their multiplicity ratio 5/3.
Errors are mostly due to statistics, but also cover
variations of centrifugal barriers and form factors discussed below.
The first three components of Table 1 are very stable in
intensity, whatever changes are made to other components;
they are also insensitive to the choice between a Breit-Wigner amplitude
or a Flatt\' e form in fitting $f_2(1565)$ and to its precise mass and width.
This insensitivity reflects the fact that the $\omega \omega$ correlations are
able to identify the spin dependence distinctively.

\begin {table}[htp]
\begin {center}
\begin {tabular} {lc}
Channel & contribution (\%) \\\hline
$^1S_0 \to f_2(1565)\pi $ & $(4.2 \pm 0.2)$ \\
$^3P_1 \to f_2(1565)\pi $ & $(19.5 \pm 2.8)$\\
$^3P_2 \to f_2(1565)\pi $ & $(30.9 \pm 3.3)$\\
$^1S_0 \to b_1(1235)\omega + \sigma \pi$ & $(45.4 \pm 4.2)$\\\hline
\end {tabular}
\end {center}
\caption {Percentage contributions to $\omega \omega \pi ^0$ data. }
\end {table}

We discuss now the formula used to fit production of the $f_2(1565)$:
\begin {eqnarray}
f_{1565} &=& \frac {\Lambda B_L(p)V(p)\exp (-\alpha q^2)}
{M^2 - s - m(s) -iM\Gamma _{tot}}, \\
\Gamma _{tot} &=& \Gamma _{2\pi} + g_1\rho _1 + g_2\rho _2.
\end {eqnarray}
In the numerator of equn. (7), $V(p)$ is the Vandermeulen form factor for the
production process $\bar pp \to f_2(1565)\pi$ [15];
$p$ is the centre of mass momentum of $f_2(1565)$.
The factor $B_L(p)$ is the standard Blatt-Weisskopf centrifugal barrier
for production with orbital angular momentum $L$.
Elsewhere, we find a radius for the centrifugal barrier $R = 0.5$ to 1.0 fm;
here, we extend the possible range to 1.3 fm, to accomodate the
possibility that $f_2(1565)$ is an $\omega \omega$ `molecule' of large radius.
Fits to data are insensitive to $R$, but it will play a role in
unfolding the line-shape of the resonance.
The exponential is a form factor with $\alpha = 1.5$ GeV$^{-2}$ 
for decay of the resonance to $\omega \omega$, each of
momentum $q$. 
The factor $\Lambda$ is a complex coupling constant for $^1S_0$ production;
for P-state production it may be taken to be real, since only
$f_2(1565)$ contributes.
Other processes are described by equations analogous to (7), except that
$B_L(p) = 1$ and $\Gamma _{tot}$ is constant.

\begin{figure}
\begin{center}
\epsfig{file=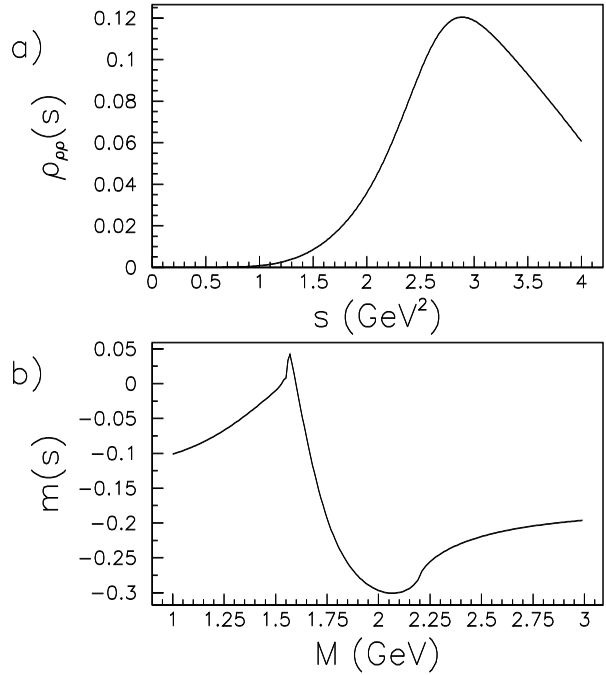,width=8cm}\
\vskip -10.02cm
\epsfig{file=RHOANDM2.EPS,width=8cm}\
\vskip -5mm
\caption{(a) The phase space factor for $\rho \rho$ as a function of 
$s$; (b) $m(s)$.}

\end{center}
\end{figure}

In equn. (8), the $2\pi$ width is taken to be constant.
We shall find that it is very small.
Care is therefore necessary over the $s$-dependence of the widths for
decays to $\rho \rho$ and $\omega \omega$.
The factors $\rho _{1,2}$ of equn. (8) are phase space factors for
$\rho \rho$ (channel 1) and  $\omega \omega$ (channel 2).
For the $\omega \omega$ channel, $\rho _2= 2q\exp (-2\alpha q^2)/\sqrt {s}$.
For the $\rho \rho$ channel, the $\rho _1(s)$ is evaluated numerically
from equn. (40) of Bugg, Sarantsev and Zou [16].
It is illustrated in Fig. 4.
It has been parametrised by the following empirical expression:
\begin{equation}
\rho _1(s) = \frac {-3.909+10.571s-1.81s^2}{48.47[1+\exp(11.353(1.063-s)
+s^2(4.572-0.826s))]},
\end{equation}
with $s$ is GeV.
The quark model predicts $g_1 = 3g_2$, corresponding to the three charge
states of $\rho \rho$ and one for $\omega \omega$.
We explore other values of the ratio
$g_1/g_2 = 1$ to 4, because of the possibility that $f_2(1565)$
is a `molecule'.

In equn. (7), $m(s)$ is an important dispersive correction to the mass
evaluated from the subtracted dispersion relation:
\begin {equation}
m(s) = \frac {(s - M^2)}{\pi }\int \frac {ds'~M\Gamma _{tot}(s')}{(s' - s)
(s' - M^2)}.
\end{equation}
Here, $\Gamma _{tot}$ is the total width appearing in equn. (8). 
This dispersive correction makes the amplitude fully analytic.
It has been checked by evaluating the standard dispersion relation
for the real part of the amplitude in terms of an integral over its
imaginary part; this relation is accurately satisfied over the
whole range of $s$ relevant here.

Our objective is now to make a combined fit to the $\omega \omega \pi ^0$
data and $3\pi ^0$ to determine $M$,
$\Gamma _{2\pi }$, $g_1$ and $g_2$.
We also use VES $\omega \omega$ data to constrain the width of the
$\omega \omega$ peak at half-height.
In practice, what we find is as follows.
Firstly, the $\omega \omega \pi ^0$ data depend fairly weakly on the
mass of the resonance and its full width.
Their main role is to determine the magnitudes of the $^1S_0$, $^3P_1$
and $^3P_2$ components. Secondly,
the $3\pi ^0$ and $\omega \omega \pi ^0$ data cannot be
fitted simultaneously with a simple Breit-Wigner amplitude of
constant width; this would require differences between channels
for mass and width outside errors.
They are, however, readily fitted together by equns. (7) and (8).

The fit to VES data and $3\pi ^0$ determines well the mass and the
full width of the resonance $\Gamma _{tot}$.
The $3\pi ^0$ data define the lower side of the resonance and VES data the
upper side. 
The fit to $3\pi ^0$ uses the same ingredients as Abele et al. [6].
It is sensitive to the $^1S_0$ component of production of
$f_2(1565)$, since this component interferes with other strong
amplitudes. It is insensitive to the magnitudes of the
$^3P_2$ and $^3P_1$ amplitudes for producing $f_2(1565)$, since they
interfere only with the weak $f_2(1270)\pi$ channel;
the intensities of $^3P_2$, $^3P_1 \to f_2(1565)$ are small 
($\sim 1.2$ and $0.7\%$ respectively).
[They are smaller with respect to $^1S_0$ than in Table 1 because
the $\pi \pi$ data extend to lower masses than $\omega \omega$
and the centrifugal barriers therefore create less suppression.]

The value of $M$ optimises at 1598 MeV, with a statistical error of $\pm 11$
MeV.
Systematic variations with $R$ and $g_1/g_2$ are $\pm 9$ MeV.
The dispersive correction $m(s)$ plays a vital role in this stability.
As $M$ is varied, the subtraction point in the dispersion integral
alters and $m(s)$ moves up and down bodily at all $s$; the data
will not tolerate much movement of this term, which is illustrated in
Fig. 4(b).
Without $m(s)$, the quality of the combined fit deteriorates sharply.
A detail is that the narrow spike in $m(s)$ between 1.56 and 1.6 GeV
originates from the rapid opening of the $\omega \omega$ channel.

The full width at half maximum of the summed cross section for
$\pi \pi$, $\omega \omega$ and $\rho \rho$ channels is determined reliably as 
 $220 \pm 15$ MeV.
For $g_1/g_2 = 3$, this corresponds to $g_1 = 435 \pm 30$ MeV,
with a systematic error from variations of $R$ and
$g_1/g_2$ which is negligible compared with the statistical error.
However, the split between $\Gamma _{\rho \rho }$ and $\Gamma _{\omega \omega
}$ is uncertain.

The value of $\Gamma _{2\pi }$ is very small: $2.4$ MeV.
This small value follows from the very small intensity for
production of $f_2(1565)$ in $\bar pp \to 3\pi ^0$ from $^1S_0$.
This is $(0.50 \pm 0.05)\%$ of the $3\pi ^0$ channel;
using the branching fraction $(6.2 \pm 1.0) \times 10^{-3}$ for
$\bar pp \to 3\pi ^0$ [17] and allowing for 3 charge states, the
resulting branching ratio for production of $f_2(1565)$ in $\pi \pi$
decays is $(0.93 \pm 0.18)\times 10 ^{-4}$.
The rate for production of $f_2(1565)$ from $^1S_0$ and decay to
$\omega \omega$ is somewhat higher:
$0.047 \times 4.5 \times  10^{-3} = 2.2 \times 10^{-4}$.
However, this latter figure hides the fact that production of $f_2(1565)$ 
from $^1S_0$ is highly suppressed by the centrifugal barrier at the high 
masses available for $\omega \omega$ decay.
To quantify this, we unfold the effect of the centrifugal barrier, 
and also the obvious suppression due to the limited phase space available 
in $\bar pp$ annihilation;
the latter suppression factor is $p/M$, where $p$ is the centre of mass
momentum of the $f_2(1565)$ and $M$ its mass.
For $g_1/g_2 = 3$,
we find for $\Gamma _{2\pi }$,
$\Gamma _{\omega \omega}$ and $\Gamma _{\rho \rho }$, averaged over the
line-shape of the resonance:
\begin {equation}
\bar {\Gamma }_{2\pi }:\bar {\Gamma }_{\omega \omega }:
\bar {\Gamma }_{\rho \rho } = 0.06^{+0.12}_{-0.04}:1.23 \pm 0.6:3.
\end {equation}
The errors on $\bar {\Gamma }_{2\pi }$ express a factor 3 uncertainty in 
$\bar {\Gamma }_{2\pi }/\bar {\Gamma }_{\rho \rho }$ arising from 
uncertainties in the centrifugal barrier.
There is also a factor 1.5 uncertainty in
$\bar {\Gamma }_{\omega \omega }$ from this source.
Despite these sizeable errors, what is clear is that the $2\pi$ width is 
very small.
This explains why the $f_2(1565)$ has not been observed in $\pi \pi \to
\pi \pi$, e.g. in the CERN-Munich experiment [18].

In equn. (11), the ratio of $\rho \rho$ widths to $\omega \omega$ is
of course dependent on $g_1/g_2$.
What is accurately determined is the full width.
The value of $g_1/g_2$ cannot be significantly less than 2, otherwise the
$\pi \pi$ channel is too strong at low masses and the fit to the $3\pi ^0$
data deteriorates sharply.
A small detail is that, in equn. (11), $ \bar {\Gamma }_{\omega \omega }/
\bar {\Gamma }_{\rho \rho }$ is
just above the value 3 assumed for $g_1/g_2$.
The reason is that the finite width of the two $\rho $ suppresses the
$\rho \rho$ signal slightly in the mass range of the $f_2(1565)$.

 \begin{figure}
\begin{center}
\vskip -20mm
\epsfig{file=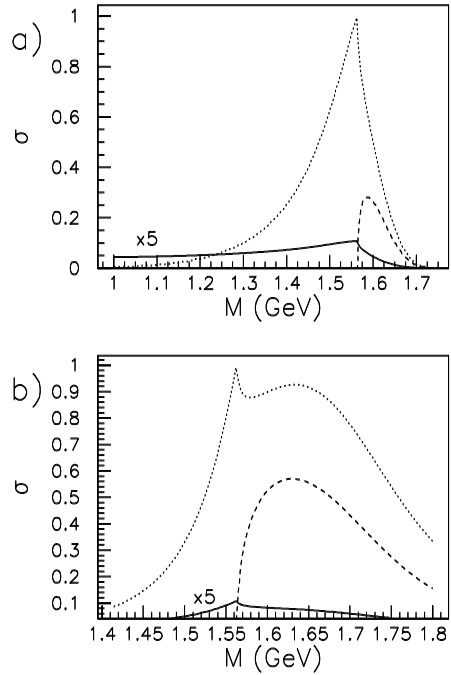,width=7cm}\
\vskip -10.55cm
\epsfig{file=PHASEE2.EPS,width=7cm}\
\vskip -6mm
\caption{The intensity distribution for $f_2(1565) \to \pi \pi$ (full curve),
$\omega \omega$ (dashed) and $\rho \rho $ (dotted) for production in $\bar pp$
annihilation from $^1S_0$; (b) as (a) for the $f_2(1565)$ uninhibited by the
centrifugal barrier effects or by phase space for production in 
$\bar pp$ annihilation.}
 \label{ph1.eps}
  \end{center}
   \end{figure}

To clarify these results, we display on Fig. 5(a) the shape of the
resonance in $2\pi$, $\omega \omega$ and $\rho \rho$ channels of
production from $^1S_0$.
The $2\pi$ result is what is fitted to $3\pi ^0$ data (scaled by a factor 5
for purposes of display).
It is strongly asymmetric, falling to half-height
at 1.59 GeV because of the
rapidly opening $\omega \omega$ and $\rho \rho$ channels.
The $\omega \omega$ result is likewise what is fitted to $\omega \omega
\pi ^0$ data for $^1S_0$ annihilation.
It peaks at 1.59 GeV, but is rapidly attenuated by the centrifugal
barrier.
The $\rho \rho$ channel peaks at the $\omega \omega$ threshold;
it falls on the lower side
because of the falling phase space and on the upper side because of the
centrifugal barrier and the opening of the $\omega \omega$ channel.

Fig. 5(b) shows our estimate of corresponding results for an $f_2(1565)$
in isolation, uninhibited by the phase space for production in
$\bar pp$ annihilation or by the centrifugal barrier there.
The results are subject to a factor 3 uncertainty of scale and some
uncertainty of shape in unfolding the
effect of the centrifugal barrier.
Nonetheless, they illustrate one unavoidable feature.
There is a cusp in the $\pi \pi$ channel at the $\omega \omega$ threshold,
 clearly visible in Fig. 5(b).
The $\rho \rho$ and $\omega \omega$ channels peak at $\sim 1.63  $
and 1.66 GeV respectively.
A Breit-Wigner amplitude of constant width is a poor approximation.

In summary, we find that $f_2(1565)$ is produced in the
$\bar pp \to \omega \omega \pi ^0$ channel with well determined branching
fractions from $^1S_0$, $^3P_1$ and $^3P_2$; P-state annihilation
dominates strongly.
It is well fitted by a resonance with strongly $s$-dependent widths in
$\rho \rho$ and $\omega \omega$ channels and a mass of
$1598 \pm 11(stat) \pm 9(syst)$ MeV.
Unfolding the effects of the uncertain centrifugal
barrier is problematical, but the $2\pi$ width is certainly very small,
of order 2\% of the total.

The $f_2(1565)$ makes a natural candidate for the radial excitation of
$f_2(1270)$.
However, the fact that its binding energy in $\rho \rho$ and $\omega \omega$
channels is small implies there must be some `molecular' tail to its
wave function, analogous to the long range tail of the deuteron.

\section{Acknowledgement}

We wish to thank the technical staff of the LEAR machine group
and of all the participating institutions for their invaluable
contributions to the success of the experiment. We acknowledge financial
support from the British Particle Physics and
Astronomy Research Council,
the German Bundesministerium f\"ur Bildung, Wissenschaft,
Forschung und Technologie, the U.S.~Department of Energy and
the National Science Research Fund Committee of Hungary (contract
No.~DE-FG03-87ER40323,
DE-AC03-76SF00098, DE-FG02-87ER40315 and OTKA T023635).
N.~Djaoshvili acknowledges support from the DAAD.


\begin {thebibliography} {99}
\bibitem {1} B. May et al., Phys. Lett. B225 (1989) 450.
\bibitem {2} B. May et al., Z. Phys. C46 (1990) 203.
\bibitem {3} A.  Adamo et al., Nucl. Phys. A558 (1993) 13C.
\bibitem {4} V.V. Anisovich et al., Phys. Lett. B323 (1994) 233.
\bibitem {5} C. Amsler et al., Phys. Lett. B355 (1995) 425.
\bibitem {6} A. Abele et al., Nucl. Phys. A609(1996) 562.
\bibitem {7} D. Alde et al., Phys. Lett. B241 (1990) 600.
\bibitem {8} A. Beladidze et al., Z. Phys. C54 (1992) 367.
\bibitem {9} Particle Data Group, Euro. Phys. J. 3 (1998) 1.
\bibitem {10} E. Aker et al., Nucl. Instr. and Methods A321 (1992) 69.
\bibitem {11} C. Amsler et al., Phys. Lett. 327B (1994) 425.
\bibitem {12} C. Hodd, Ph. D. thesis, University of London (1999).
\bibitem {13} D.V. Bugg et al., Phys. Lett. B353 (1995) 378.
\bibitem {14} U. Thoma, AIP Conf. Proc. 432 (1998) 322.
\bibitem {15} J. Vandermeulen, Z. Phys. C37 (1988) 563.
\bibitem {16} D.V. Bugg, A.V. Sarantsev and B.S. Zou, Nucl. Phys. B471 (1996)
59.
\bibitem {17} C. Amsler et al., Phys. Lett. B342 (1995) 433.
\bibitem {18} B. Hyams et al., Nucl. Phys. B64 (1973) 134.
\end {thebibliography}
\end {document}